\documentstyle{l-aa}
\input epsf.sty

\begin{document}

   \thesaurus{12         % A&A Section 12: 
              (02.01.2;  % Accretion discs
               02.18.8;  % Relativity
               02.18.5)} % Radiation mechanisms: non-thermal

  \title{
   Optics near Kerr black holes: 
    spectra of advection dominated accretion flows.
          }

   \subtitle{}

   \author{ M. Jaroszy\'nski 
           \and A. Kurpiewski }

\institute{Warsaw University Observatory, Al. Ujazdowskie 4,
 00--478 Warsaw, Poland
}

%\date{Received 11 March 1997; Accepted 25 March 1997}

\maketitle

\begin{abstract}

We investigate advection dominated, trans\-sonic accre\-tion flows
in the vicinity of a Kerr black hole. We take into account all
relativistic effects in the dynamics of the flow and in the
propagation of light. We assume the matter to be weakly magnetized
and cool via the thermal synchrotron and Bremsstrahlung radiation.
We include also the effects of Comptonization. We calculate the
spectra of radiation as seen by observers located at different
positions relative to the equatorial plane of the disk. The radiation
emitted by the accreting matter is anisotropic and observers near
the equatorial plane register a higher energy flux. This effect
is more pronounced in the case of slowly rotating black holes.
We calculate also the shape of a hypothetical gamma line, which
may be produced by the thermonuclear reactions in the inner part
of the flow. The line is strongly broadened, but the fact that
the flow is quasi-spherical removes the two-peak shape of the
line seen in the spectra emitted from thin, Keplerian disks.
The kinematics of the advection dominated flows is not unique
(as opposed to Keplerian disks or spherical free-fall) and it
would probably be difficult to find strong limits on source
models using the spectral observations.

\keywords{accretion processes  - radiation mechanisms: non-thermal 
-relativity }

   \end{abstract}

%
%________________________________________________________________

\section{Introduction}

The advection dominated accretion flows (ADAFs) have recently 
become an interesting alternative to the traditional thin 
accretion disks (Shakura \& Sunyaev 1973; Novikov \& Thorne 1973).
The disk model with a two temperature, optically thin and hot inner
region has been proposed by Shapiro, Lightman \& Eardley (1976). 
The spherical accretion flow, which cannot cool, because the radiation is
trapped and advected with matter, has been considered by Katz (1977).
Spherical accretion with advection has been
investigated by Begelman (1978, 1979). The role of advection in
disk-like accretion has been studied by Begelman \& Meier (1982),
Abramowicz, Lasota \& Xu (1986) and Abramowicz et al. (1988) 
for configurations with substantial optical and geometrical thickness.
These models are sometimes called {\it slim accretion disks}.
The advection of heat makes the energy conservation nonlocal, as opposed
to thin disks, and thus influences their structure, but due to the 
substantial optical thickness, has no prominent influence on their 
spectrum. Another branch of accretion flows has been discovered 
by Narayan \& Yi (1995) and Abramowicz et al. (1995). Flows belonging 
to this class are extremely optically thin and have low accretion rate.
Because of the low radiation efficiency, the gas can heat up to near
virial temperatures. For this reason the flow becomes quasi-spherical
and pressure gradients play an important role in the gas dynamics. The
radial and azimuthal velocities can have the same order of magnitude.
These are 
the main characteristics of ADAFs. Because of their high temperatures
ADAFs produce X-rays in a natural way. On the other hand the 
effectiveness of accretion is low, because almost all the thermal energy
goes under the horizon with accreting matter. 
These properties make ADAFs possible explanation for some X-ray sources
and such models have been proposed (Narayan 1996; Lasota et al. 1996;
Narayan, McClintock \& Yi 1996; see also Rees, 1982).

The general properties of the advective flows have been obtained 
with the help of Newtonian or pseudo-newtonian (Paczy{\'n}ski \& Wiita
1980) description of the
background gravitational field. Lasota (1994), Abramowicz et al. (1996,
hereafter ACGL) and Abramowicz, Lanza \& Percival (1997, hereafter ALP) 
have obtained
a self-consistent system of equations describing the accretion flows in
the Kerr spacetime. All these papers use the {\it vertical averaging} 
to obtain a set of ordinary differential equations  
to describe the flow. The form of equations resembles the equations for 
an accretion disk, but some properties of the quasi-sphericity 
are taken into account (see ALP). The properties of ADAFs in fully 
relativistic treatment have been also investigated by Peitz \& Appl (1996).
All these papers investigate mainly the flow dynamics and are concentrated
on its properties and their dependence on parameters such as viscosity 
parameter $\alpha$ and the dimensionless accretion rate $\dot m$. 

We are interested in the radiation processes and spectra emitted by ADAFs
in the Kerr spacetime. The main reason of our investigation is to get
a fully self-consistent, relativistic description of the radiation from 
the advective flows. The gas has the highest temperatures near the 
horizon, so the combined effects of ray deflection, Doppler shifts 
and gravitational redshift may influence the high energy end of the 
spectrum, at least quantitatively. Since the flow is optically thin,
the observer may have a chance to see the matter on the other side 
of the black hole approaching the telescope with relativistic speed; 
in some cases this should lead to the beaming of radiation and substantial
increase of the surface brightness. The possibility and importance 
of such effects should be investigated.

Since we are mostly interested in the propagation of light,
we are not going to investigate the hydrodynamics of the flow in detail.
To get rough solutions of the hydrodynamic equations we use 
an approximate treatment, neglecting the cooling.
Thus the most time consuming part of the 
calculations is decoupled from the equations of motion. 
For a simple equation of state the structure of the flow 
(i.e. the profile of the angular momentum, radial velocity of matter
and the speed of sound) depends only
on the viscosity parameter $\alpha$ and the values of the specific 
angular momentum and energy of matter at the horizon. 
The matter density depends on the above variables and scales with 
the accretion rate. Using the same solution of the equations of motion
one is able to get models with different accretion rate.
Inspection of the solutions of the hydrodynamic equations of 
Chen, Abramowicz \& Lasota (1997), which include cooling, shows that 
dependence on the accretion rate is weak, at least in the inner part
of the flows.

In the next Section we present the system of equations, which we use to
model the accretion flow and rays tracing. We also describe the approach
to radiation processes within the flow. In Sec.3 we present the results
of calculations, showing the brightness profiles of our sources, their
spectra, and luminosity dependence on the observer inclination angle. 
The discussion follows in the last Section.

%___________________________________________________________________
\section{The model}

\subsection{The accretion flow}

We investigate the stationary flow of matter and the propagation of light
in the gravitational field of a rotating Kerr black hole 
using the Boyer-Lindquist coordinates $t$, $\phi$, $r$, $\theta$ and the 
metric components $g_{ab}(r,\theta)$ as given by Bardeen (1973).
We follow the $(-,+,+,+)$ signature convention. 
When dealing with the flow
we use the metric tensor with components evaluated at the equatorial plane
($\theta=\pi/2$). We use geometrical units, so the speed of light
$c \equiv 1$ and the mass of the hole $M \equiv 1$. (Later we use $c$
for the speed of sound.)  The Kerr parameter $a$ ($0 \le a <1$) gives
the black hole angular momentum in geometrical units. We use the
Einstein summation convention, where needed and a semicolon for the
covariant derivative. The normalization of four velocity with the chosen
metric signature reads $u^a u_a=-1$.

The system of equations we use follows in general ACGL. Since we 
are neglecting the cooling processes at this stage, and treat the 
accretion flow in the disk approximation, two components of velocity
and the speed of sound $c$ given as functions of radius, fully describe
the dynamics. We use the angular velocity $\Omega$ and the physical
radial velocity $V$ as measured by locally nonrotating observers as
main kinematic variables (compare ACGL).
After the {\it vertical averaging} the velocity perpendicular to the 
equatorial plane is neglected ($u^\theta \equiv 0$)   and other
components of the four velocity are given as:
\begin{equation}
u^t={1 \over \sqrt{1-V^2} 
\sqrt{-g_{tt}-2 \Omega g_{t\phi}-\Omega^2 g_{\phi\phi}}}
\end{equation}
\begin{equation}
u^\phi=\Omega u^t
\end{equation}
\begin{equation}
u^r={1 \over \sqrt{g_{rr}}}{V \over \sqrt{1-V^2}}~~.
\end{equation}

We shall later use
the two temperature plasma to describe the cooling processes. In the hottest 
and most interesting part of the flow the pressure of the nonrelativistic 
ions, $p=n_i kT_i$, where $n_i$ is ion concentration, $T_i$ their temperature
and $k$ the Boltzmann constant, dominates. The isothermal sound speed
can be defined as
\begin{equation}
c^2={kT_i \over m}
\end{equation}
where $m$ is the average ion mass, so one has $p=\rho_0c^2$, where
$\rho_0$ is the rest mass density. The specific enthalpy $\mu$ is given as:
\begin{equation}
\mu \equiv {\epsilon+p \over \rho_0}=1+{5 \over 2}c^2
\end{equation}
where $\epsilon$ is the total (rest mass plus thermal) energy density.
The above formula implicitly assumes that the matter can be treated
as the ideal, nonrelativistic gas. 

The mass conservation equation relates the mass accretion rate ${\dot M}$,
with the radial velocity and the surface rest mass density $\Sigma$
(ACGL): 
\begin{equation}
{\dot M} = -2\pi\Sigma \sqrt{\Delta} {V \over \sqrt{1-V^2}}
\end{equation}
where $\Delta (r) \equiv r^2-2r+a^2$ is one of the functions defining the
metric. On the horizon one has $\Delta(r_h)=0$ and $V=-1$ (see ACGL), but the
ratio $\Delta/(1-V^2)$ remains finite and is related to the derivative
of radial velocity at $r=r_h$. Thus the surface mass density remains
finite everywhere and using mass conservation one can substitute for
$\Sigma$ in other equations. 

The rest mass density of matter can also be replaced in the equations
with the help of the so called {\it vertical hydrostatic equilibrium
equation}, which reads:
\begin{equation}
{dp \over dz} = -\mu \rho_0 g_{,z} z
\end{equation}
where $g_{,z}$ represents the gravitational tidal forces as measured at
the equatorial plane by an observer comoving with the fluid, and $z$ is
his coordinate in this direction. 
We use the prescription of ALP for the tidal forces, so the expression
remains finite at the horizon:
\begin{equation}
g_{,z}={u_\phi^2-a^2(u_t^2-1) \over r^4}
\equiv {\ell_*^2 \over r^4}
\end{equation}
For an optically thin medium we assume the temperature to be
approximately constant in the vertical direction. 
The integration of of the equation gives the density distribution:
\begin{equation}
\rho_0(r,z)=\rho_0(r,0)\exp{\left(-{1 \over 2}{z^2 \over H^2}\right)}
\end{equation}
where
\begin{equation}
H^2 = {c^2 \over \mu}{r^4 \over \ell_*^2}
\end{equation}
is the scale height. 
The form of the above expression shows, that $c^2/\mu$ has the physical
meaning of the square of the speed of sound.
Integration of the density distribution
over the disk thickness gives $\Sigma = \sqrt{2\pi} \rho_0 H$. Knowing
the surface mass density and the scale height one can substitute for the
rest mass density.

We are neglecting transport of heat within the fluid, but we do include
the processes due to the shear viscosity, so the energy momentum tensor
has the form (Novikov \& Thorne 1973):
\begin{equation}
T^a_b=(\epsilon + p)u^a u_b + p \delta^a_b -2 \eta \sigma^a_b
\end{equation}
where $\delta^a_b$ is the Kronecker delta, $\eta$ the dynamical
viscosity, and $\sigma^a_b$ the shear tensor:
\begin{equation}
\sigma_{ab}={1 \over 2} (u_{a;i}h^i_b+u_{b;i}h^i_a)
            -{1 \over 2}\Theta h_{ab}
\end{equation}
The equations of motion for the disk are written under the assumption
that the velocity in the vertical direction can be neglected and all
variables are evaluated in the equatorial plane. The motion is
effectively two dimensional and all tensors should be treated as acting
in  three dimensional space time. The projection tensor 
$h^a_b=\delta^a_b+u^a u_b$ and its $3^d$ trace $h^a_a=2$. This is why we
use the factor $1/2$ instead of conventional $1/3$ in front of the term
including the velocity expansion $\Theta=u^a_{;a}$ in the definition 
of the shear tensor.

The projected equation of motion, $h_r^bT^a_{b;a}=0$ takes the form:
$$
{V^2-c^2/\mu \over 1-V^2}{V^\prime \over V}
  +2{c^2 \over \mu}{c^\prime \over c}=
{1 \over 2}{c^2 \over \mu}{\Delta^\prime \over \Delta}
~~~~~~~~~~~~~~~~~~~~~~~~~~~~~~~~~~~~~
$$
\begin{equation}
~~~~~~~~+{1 \over 2}(1-V^2)(u^t)^2
(g_{tt}^\prime+2\Omega g_{t\phi}^\prime+\Omega^2 g_{\phi\phi}^\prime)
\end{equation}
where the primes denote the derivatives relative to $r$. 
We neglect the terms related to the viscosity in this equation, which is
the usual simplification (ACGL, Peitz \& Appl 1997).

Projecting the equation of motion into the comoving reference frame,
$u^b T^a_{b;a}=0$ leads to the energy conservation equation in the form
(Novikov \& Thorne 1973, ACGL):
\begin{equation}
\rho_0 T u^a \nabla_a S = 2\eta \sigma^2
\end{equation}
where $S$ is the entropy of the fluid per unit mass and the derivative is
in fact relative to the proper time. In the stationary, axially
symmetric case $u^a \nabla_a =u^r \partial_r$. The entropy of an ideal
gas is given as:
\begin{equation}
S={k \over m} ({3 \over 2}\ln T - \ln \rho_0) +const
\end{equation}
For the viscosity in the {\it $\alpha$-disk} approximation one has:
\begin{equation}
\eta = \alpha \rho_0 {c \over \sqrt{\mu}} H 
=\alpha \rho_0 {c^2 \over \mu} {r^2 \over \ell_*}
\end{equation}
where we use the expression for the sound speed with the proper physical
meaning. 
The equation for the entropy gradient can now be expressed as:
\begin{equation}
{V^\prime \over V}+3{c^\prime \over c}+{H^\prime \over H}=
-{1 \over 2}{\Delta^\prime \over \Delta}
-{2\alpha r^3 \over \mu \ell_*}\sqrt{(1-V^2) \over \Delta V^2}~~\sigma^2
\end{equation}
where the radial derivative of the scale height is still to be expressed
via the gradients of the angular and radial velocities and the speed of
sound. 

The symmetries of the Kerr metric imply the energy and angular momentum
conservation equations for the fluid in the form $T^a_{t;a}=0$,
$T^a_{\phi;a}=0$. After integrating over r and vertically averaging one
has:
\begin{equation}
{\dot M \over 2\pi}(\mu u_\phi-\mu^0u^0_\phi)
+2\alpha r \Sigma{c^2r^2 \over \mu \ell_*}\sigma^r_\phi=0
\end{equation}
\begin{equation}
{\dot M \over 2\pi}(\mu u_t-\mu^0u^0_t)
+2\alpha r \Sigma{c^2r^2 \over \mu \ell_*}\sigma^r_t=0
\end{equation}
where the superscript $0$ denotes variables as measured on the horizon.
We define the specific energy of the fluid as ${\cal E} = -\mu u_t$
and its specific angular momentum as ${\cal L}=\mu u_\phi$. We also
use the kinematic specific angular momentum $\ell=-u_\phi/u_t$.
Combining the upper equation multiplied by ${\cal E}$ with the lower
equation multiplied by ${\cal L}$ we get:
\begin{equation}
{\alpha c^2r^3 \over \ell_*}(2\sigma^r_t u_\phi-2\sigma^r_\phi u_t)=
-\sqrt{\Delta V^2 \over 1-V^2}{\cal E}{\cal E}_0(\ell - \ell_0)
\end{equation}
The combination of the shear tensor components, which we use, contains the
angular velocity gradient but not the radial velocity gradient:
$$
2\sigma^r_t u_\phi-2\sigma^r_\phi u_t=
~~~~~~~~~~~~~~~~~~~~~~~~~~~~~~~~~~~~~~~~~~~~~~~~~~~~~~~~~~~~
$$
\begin{equation}
~~~~{\Delta (u^t)^2 \over 1-V^2}\Omega^\prime
+{V^2 [\ell
g_{tt}^\prime+(1+\Omega\ell)g_{t\phi}^\prime+\Omega g_{\phi\phi}^\prime]
\over (1-V^2)^2 (1-\Omega\ell)}
\end{equation}
Thus the equation for $\Omega $  is obtained. It has a boundary condition 
at the horizon $\Omega=\omega$, where $\omega=-g_{0\phi}/g_{\phi\phi}$ 
is the angular velocity of the dragging of inertial frames.

We solve the equations for $V$, $c$ and $\Omega$ by iterations. We start
from a solution with subkeplerian rotation, which asymptotically (far
from the hole) becomes
a selfsimilar solution as described by Narayan \& Yi (1994). We
postulate the position of the sonic radius $r_s$ and the value of the
angular momentum at the horizon $\ell_0$ and, using the angular velocity
distribution from the initial guess (and later on from the previous
iteration), we solve the equations for the radial velocity and the speed of
sound. These equations are easy to integrate if one starts from the
sonic point. For a given angular velocity distribution, in particular 
knowing its value at the sonic point $\Omega_s$, we find the values 
of the other two variables $V_s$, $c_s$ from the regularity conditions.
The boundary condition on the horizon ($V=-1$) is always
met. The value of specific energy on the horizon results from the
solution and is iterated. Next we solve the equation for the angular
velocity starting from the horizon outward, which gives a new angular
momentum distribution. The expression for the square of the shear 
($\sigma^2$) contains derivatives of the angular and radial velocities,
and is rather complicated, including the derivatives in the second power.
Using Eq.(12) we calculate the shear tensor from the 
``previous iteration'' and substitute it into Eq.(17). 
 
Not all postulated ($r_s$, $\ell_0$) pairs lead
to solutions which can be continued from the horizon to infinity, but
the iterations just described are so fast that checking many of them is
easy. For a given angular momentum of the black hole we find the {\it
best} parameters, for which the solution can be smoothly joined with the
selfsimilar solution far away.

\begin{figure}
%\picplace{13.0cm}
%%\plotfiddle{F1.ps}{16.0cm}{0}{50}{-160}{90}
\epsfxsize = 75 mm \epsfysize = 130 mm
\epsfbox[160 350 380 720] {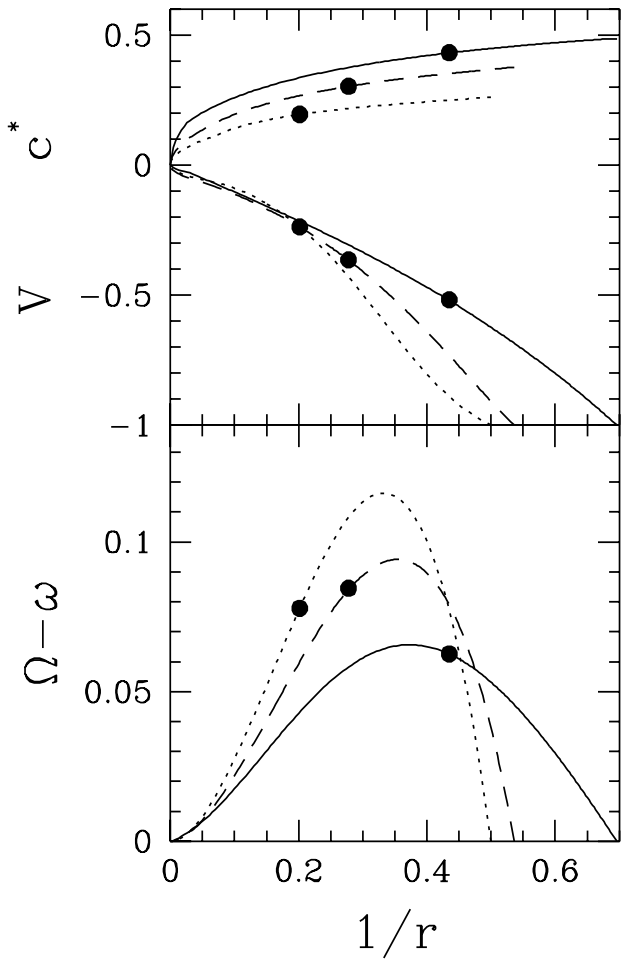}
\caption[]{Sample models of ADAFs. On the upper plot we show the
radial velocity of the flow $V$ (negative) and the sound speed
$c^* \equiv c/\sqrt{\mu}$ as functions of $1/r$ for our three models 
in the Kerr metric
with angular momentum $a=0$ (dotted line), $a=0.5$ (dashed line) and 
$a=0.9$ (solid line). The locations of the sonic points are shown as
large dots on the lines. On the lower plot we show the angular velocity
distribution $\Omega-\omega$, using the same conventions.
}
\end{figure}

For the purpose of investigating the emitted spectra we choose examples
of ADAF solutions in Kerr metric with angular momentum $a=0$, $0.5$ and
$0.9$. We use the viscosity parameter $\alpha=0.1$. For the
Schwarzschild case the position of sonic radius and the value of the
angular momentum on the horizon are in good agreement with Chen et al.
(1997), who use the pseudo-Newtonian potential to describe gravity. The
profiles of the sound speed and radial and angular velocities for our
solutions are shown on Fig.1.

\subsection{Radiation processes in the disk}

We use two temperature plasma to describe radiation processes and local
spectra in the disk. For given ion tem\-pera\-ture $T_{i}$, scale height
$H$ and surface mass density $\Sigma$ (Section 2.1.) we determine
electron temperature by taking into account the detailed balance of
their heating and cooling. We divide the disk into the coaxial cylindrical 
layers 
and treat them separately. We imagine each of the annuli to be cut from 
a plane parallel layer with constant temperature and the density 
concentrated toward the midplane according to Eq(9).
We follow here the description of heating and
cooling processes presented by Narayan \& Yi (1995) and 
also by Esin et al.
(1996) [and references therein]. They consider the energy transfer from
ions to electrons via Coulomb collisions and cooling of electrons by
synchrotron radiation, inverse Compton process and bremsstrahlung
emission. They ignore the presence of $e^+e^-$ pairs but as
Bj{\"o}rnsson et al. (1996) showed, the role of $e^+e^-$ pairs in the
structure of ADAF is not significant.

For calculated electron temperature we can model the local continuum
emission per unit area from surface of each annulus. We
assume that molecular weights of ions and electrons are $\mu_i = 1.29$
and $\mu_e = 1.18$, and the density of magnetic field is parameterized by
the ratio of gas pressure to total pressure $\beta$. In our
calculations we ignore the radiation pressure, so the total pressure is
the sum of gas pressure and magnetic field pressure. We use the
approximate formulae of Esin et al. (1996) to calculate synchrotron
spectrum and formulae of Jones (1968) to model the inverse Compton
scattering of synchrotron photons. The method of approximate computation
of bremsstrahlung spectrum is given in Svensson (1982) [and references
therein]. We include the frequency averaged comptonization of
bremsstrahlung photons in the energy balance, but we neglect this 
process in the calculation of the spectrum, since it is very time consuming
and its contribution to the resultant spectrum is not significant. 

Figure 2 shows the continuum emission calculated for four annuli of
radii $r/r_g = 1.5; 10; 100; 1000$, (where $r_g=GM/c^2$ is the 
gravitational radius)
and for the following parameters: $M/M_{\odot} = 3.6\cdot 10^7$,
$\dot{M}/{\dot{M}}_{Edd} = 0.016$, 
corresponding to 
$1.5\cdot 10^{-3} {\rm M}_{\odot}{\rm yr}^{-1}$,
$\alpha = 0.1$ and $\beta = 0.95$.
Such parameters have been chosen by Lasota et al. (1996) to model
the spectral features of the nucleus of active galaxy NGC 4258, which they
treat as a superposition of ADAF and a standard accretion disk. We show
the hottest of our models, which has $a=0.9$.

\begin{figure}
%\picplace{8.8cm}
%%\plotfiddle{F2.ps}{8.8cm}{0}{50}{-160}{90}
%\epsfxsize = 75 mm \epsfysize = 75 mm \epsfbox[10 180 500 700]{fig2.ps}
\epsfxsize = 75 mm \epsfysize = 75 mm
\epsfbox[170 330 380 540] {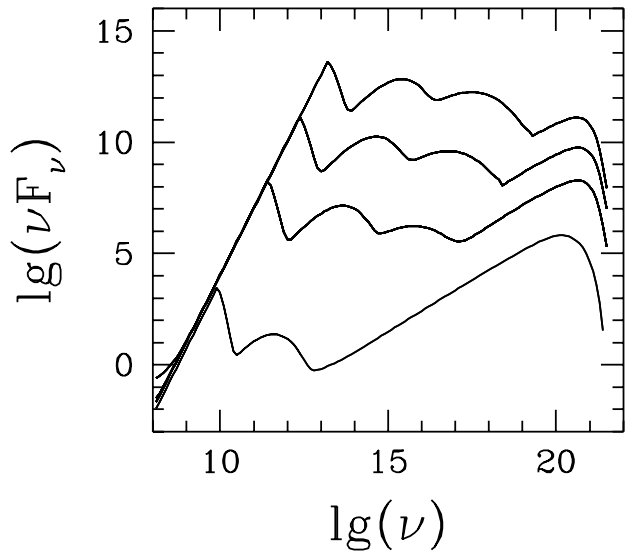}
\caption[]{Continuum emission per unit area from four annuli of radii
$r/r_g = 1.5; 10; 100; 1000$ (reading from highest line down) for
parameters: $M/M_{\odot} = 3.6\cdot 10^7$,  
$\dot{M}/{\dot{M}}_{Edd} = 0.016$, 
corresponding to 
$1.5\cdot 10^{-3} {\rm M}_{\odot}{\rm yr}^{-1}$,
 $\alpha = 0.1$, $\beta = 0.95$, and
$a=0.9$. 
The left peak of each curve is due to synchrotron radiation, the right
one is due to bremsstrahlung emission and two peaks between them are due
to the Compton scattering of synchrotron photons.}
\end{figure}

The importance of the Compton scattering depends on the total optical
depth of the disk and, strictly speaking, is not a local process. In the
following part we neglect this fact. The bremsstrahlung emission is a
two body process and, under the assumption of constant temperature in
vertical direction, we assume its rate to be proportional to the square of
the density ($\sim \rho^2$). Similarly the synchrotron emission, being
proportional to the product of electron density and magnetic pressure,
which has a constant ratio to the gas pressure and density, is also
proportional to the square of the electron concentration. The rate of
the Compton process depends on the electron concentration and the
density of the radiation field. The latter one depends on the optical
depth for scattering measured to the disk surface. 
Thus for this process the
proportionality to the square of the matter density does not hold, but
the decline toward the surface is also present and should have similar
character. We use the following, simplified formula for the volume
emissivity from the disk:
\begin{equation}
j(\nu)=\sqrt{2 \over \pi}~{F(\nu) \over \pi H}
~\exp\left(-{z^2 \over H^2}\right)
\end{equation}
The flux here is the total energy emitted from the unit surface of the
disk as usually used in this context. The emissivity $j(\nu)$ is defined
per one steradian and hence the integration of the above formula from
the equatorial plane to infinity would give $F(\nu)/\pi$.

Our formalism described in the next subsection allows for the
calculation of the source spectrum regardless of the nature of the
radiation processes in the disk as long as the matter remains
transparent to the radiation. We consider also the shape of the emission
line, which may be used for testing the velocity field in the flow.
We follow the idea of Narayan \& Yi (1996), who consider the spallation
process leading to the lithium and beryllium production. As shown by
Kozlovsky \& Ramaty (1974) some of the nuclei are produced in the excited
state and emit $\gamma$-ray lines at 478 and 431 keV. The emissivity is
proportional to the rate of production. Following closely Narayan \& Yi
(1996) we assume that there is a threshold energy $8.5$ MeV for the 
$\alpha$-$\alpha$ reaction to occur and we assume the crossection to be
inversely proportional to the energy of interacting particles
$\sigma \sim E^{-2} \sim c^{-4}$ (for $c \ge c_{\rm threshold}$). We use
here the typical thermal energy of the $\alpha$ particles instead of
energy distribution averaging. The relative velocity of interacting
particles is proportional to the thermal energy, and their concentration
to the matter density. The width of the line results from the
thermal motion of $^7{\rm Li}$ ions; they are slower than other particles
because of the higher mass, and including the $\sqrt{2}$ factor in the
definition of the line width, we get:
\begin{equation}
{\Delta \nu \over \nu_0} = 1.1c
\end{equation}
where $\nu_0$ is the line frequency and the mean ion mass $1.29m_H$ 
of the matter containing 70\% of hydrogen has been
assumed in the calculation.
Taking all the factors into account we obtain
the formula for the line volume emissivity up to a constant factor:
\begin{equation}
j(\nu) \sim {\rho^2 \over c^3}
{\exp\left(-{(\nu-\nu_0)^2 \over (\Delta \nu)^2}\right) \over \Delta \nu}
\end{equation}
where the density is obtained from Eq.(9), and the formula is valid only
for the sound velocity over the threshold.

\begin{figure}
%\picplace{16.6cm}
%%\plotfiddle{F3.ps}{8.8cm}{0}{50}{-160}{90}
\epsfxsize = 75 mm \epsfysize = 160 mm
\epsfbox[165 242 380 715] {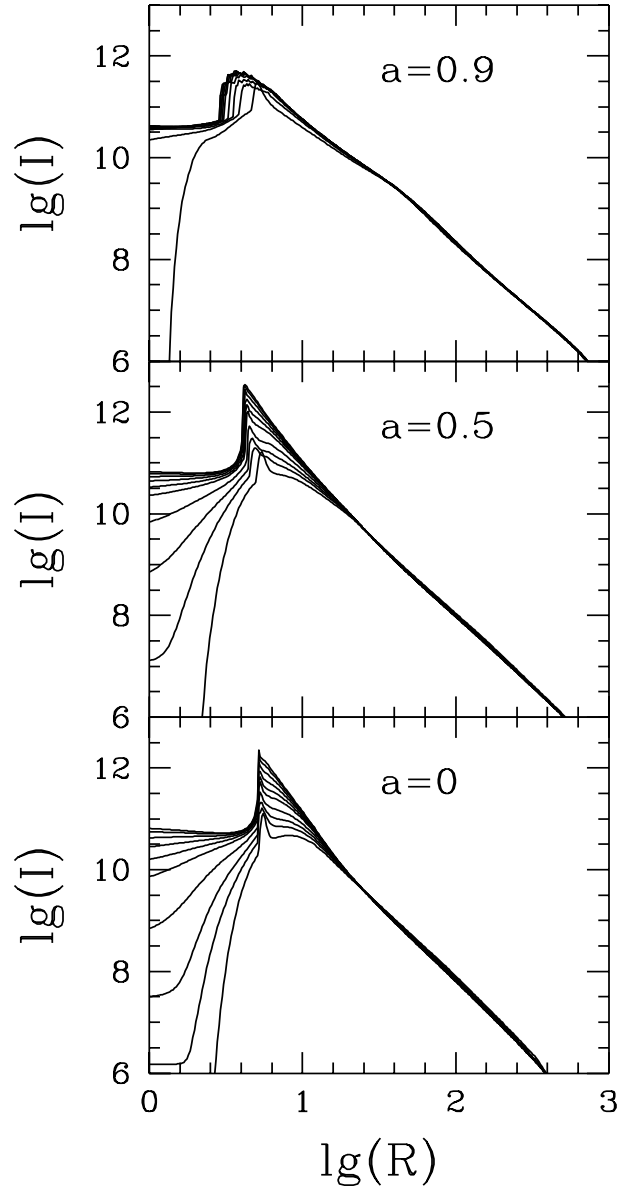}
\caption[]{Bolometric brightness profiles as seen by observers at the  
angles $\theta=5^\circ$ to $85^\circ$, bottom to top. The radius is in
geometrical units, the brightness in c.g.s. units. The three plots   
correspond to the three models as indicated.
}
\end{figure}

\subsection{Geometrical optics}

We use the backward ray shooting method (Luminet 1979; Jaroszy{\'n}ski 1993)
to obtain the spectrum of the electromagnetic radiation from the configuration.
The intensity of radiation $I(\nu_0)$ coming to a far observer 
at the frequency $\nu_0$ along any ray is given as:
\begin{equation}
I(\nu_0)=\int_{\rm ray}~~{\nu_0^3\over \nu^3}~j(\nu)~dl_{\rm prop}
\end{equation}
where $j(\nu)$ is the emissivity per unit volume, and $dl_{\rm prop}$
is the proper length differential as measured in the frame comoving with
matter. The emission frequency $\nu$ is given as:
\begin{equation}
\nu={u^a p_a \over p_0}~\nu_0
\end{equation}
where $u^a$ is the matter four velocity, $p_a$ is the photon four
momentum, and $p_0$ - the photon energy in BL frame, which is conserved. 
This is the usual redshift (or blueshift) formula.

We find the null geodesics in the Kerr metric using the quadratures of
Hamilton-Jacobi equation (Bardeen 1973). This method is faster, and more
importantly, has a better accuracy as compared with the solution of
ordinary differential equations along the trajectories.

Following many rays coming to an observer from the vicinity of the hole
makes it possible to produce a map of the surface brightness of the
source. Since the hottest part of the source is close to the hole, we
probe the radiation intensity in rings centered on the hole with
logarithmic distribution of radii. Thus the central parts are
investigated with greater accuracy. A useful characteristic of the
source is its radial surface brightness profile, as seen by a distant
observer,  obtained by integration
over frequency and azimuthally averaging:
\begin{equation}
I(\xi)={1 \over 2\pi}~
\int_0^{2\pi} d\vartheta~\int_0^{\infty} d\nu_0~I_{\nu_0}(\xi,\vartheta)
\end{equation}
where $\xi$ and $\vartheta$ are the observer cylindrical coordinates
on the sky. The length corresponding to the angle $\xi$, measured at the
source,  is $R \equiv \xi D$, where $D$ is the distance to the source.
The flux of radiation from the source as measured by an observer is
given as:
\begin{equation}
F(\nu_0) = \int d\xi~\int\xi d\vartheta~I_{\nu_0}(\xi,\vartheta)
\end{equation}
and implicitly depends on the observer location.

\section{Results}

We have made ray shooting calculations for observers at inclination
angles $\theta=5^\circ$, $30^\circ$, $40^\circ$, $50^\circ$, $60^\circ$,
 $65^\circ$, $70^\circ$, $75^\circ$, $80^\circ$ and $85^\circ$ as measured
relative to the rotation axis, for the three models of ADAF in three
background metrics with the Kerr parameter values $a=0$, $0.5$ and $0.9$.
In ray tracing we include only the radiation from a sphere of the radius
$r=10^3r_g$. We do not include any other source components, as a thin
Keplerian disk or a dust torus.

In Fig.3 we show the source surface brightness profiles as seen by
observers at different inclinations. The relatively low surface
brightness in the centre corresponds to rays, which traverse only the
matter in front of the horizon. The densest and hottest part of the
matter (close to the horizon) is receding from the observer, which also
works as to decrease the observed brightness.
The abrupt jump to higher surface brightness corresponds to rays which
pass close to the hole. These rays spend relatively long time in the
innermost part of the flow due to the bending. Some of them may be
tangent to matter trajectories in isolated points. 
Doppler boost and relative increase in optical depth (which remains much
lower than unity) both work as to increase the surface brightness.
The dependence on the inclination angle is better pronounced for the low
angular momentum black holes. The surface brightness is higher, when the 
source is watched from the vicinity of the equatorial plane.

\begin{figure}
%\picplace{8.8cm}
%%\plotfiddle{F4.ps}{8.8cm}{0}{50}{-160}{90}
\epsfxsize = 75 mm \epsfysize = 75 mm
\epsfbox[170 333 380 540] {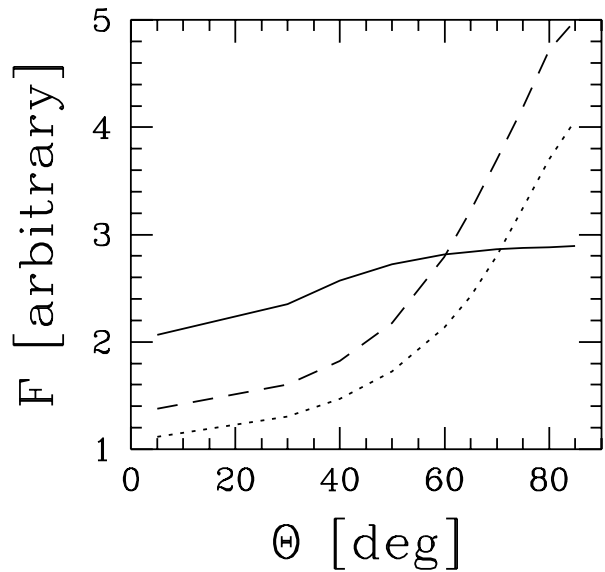}
\caption[]{The dependence of the observed total bolometric flux from the
source on the observer inclination relative to the axis of rotation.
We use the bolometric flux from our Schwarzschild model ($a=0$) 
as seen by an observer on the rotation axis as the unit of flux.
The dotted line corresponds to the nonrotating hole
($a=0$), the dashed line to the case $a=0.5$ and the solid line to $a=0.9$.
}
\end{figure}

The bolometric flux is also higher for observers close to the equatorial 
plane. As can be seen in Fig.4, the effect is moderate for the fast rotating 
hole (about 70\% increase in flux measured at the equator as compared to the
pole). For slowly rotating holes ($a=0.5$ or $a=0$) the source seen from 
the equatorial plane seems to be brighter by a factor $4$. 

\begin{figure}
%\picplace{16.6cm}
%%\plotfiddle{F5.ps}{16.6cm}{0}{50}{-160}{90}
\epsfxsize = 75 mm \epsfysize = 160 mm
\epsfbox[165 240 380 720] {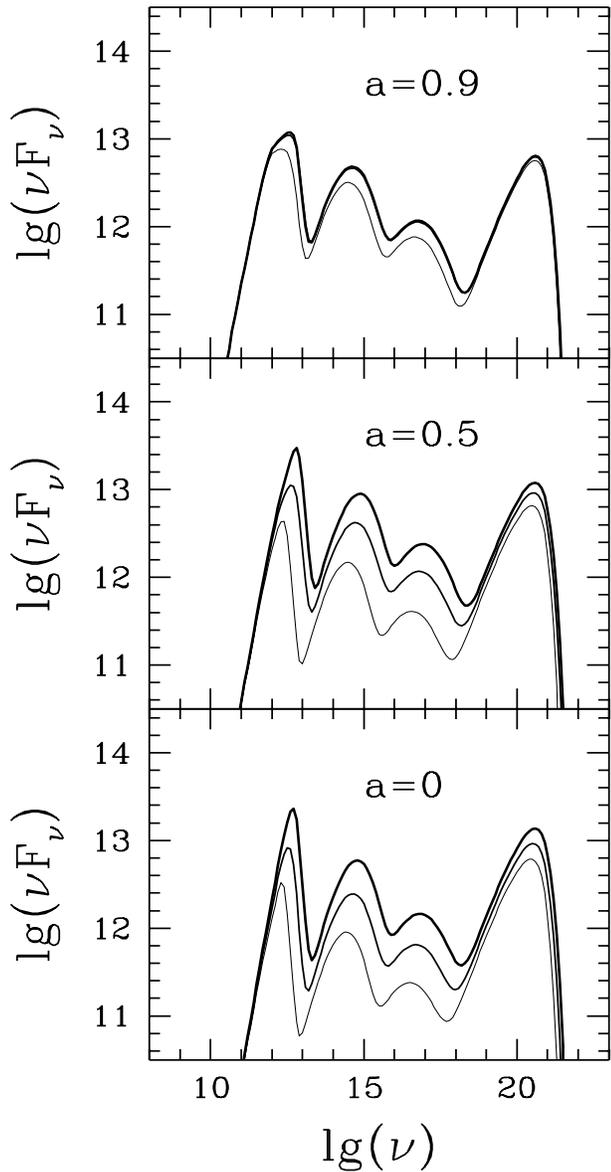}
\caption[]{The spectra of our models for observers at inclinations
$\theta=5^\circ$ (thin solid lines), $60^\circ$ (medium lines), and 
$85^\circ$ (thick lines). The three panels correspond to the three
models as indicated. The flux units are arbitrary.
}
\end{figure}

The shape of the continuum spectra from our models is not considerably
changed by the relativistic effects in the ray propagation. As seen on
Fig.5 the spectra are shifted vertically, when the observer approaches
the equatorial plane. The simultaneous shift to higher frequencies is
very weak. 

\begin{figure}
%\picplace{16.6cm}
%%\plotfiddle{F6.ps}{16.6cm}{0}{50}{-160}{90}
\epsfxsize = 75 mm \epsfysize = 160 mm
\epsfbox[165 240 380 720] {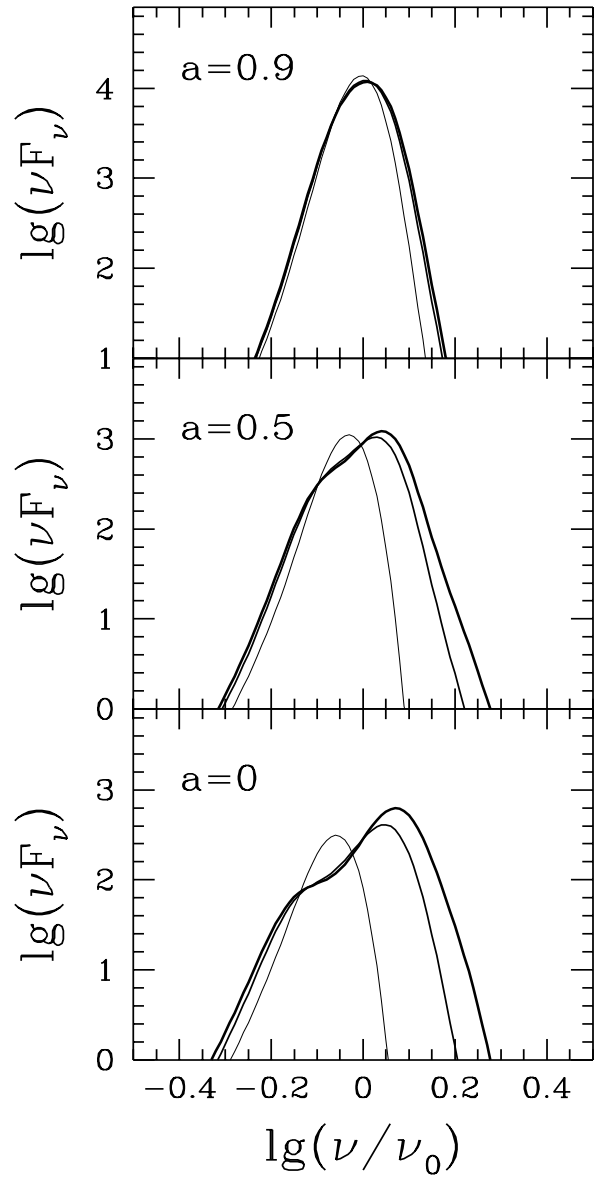}
\caption[]{The hypothetical line profiles emitted by our models
for observers at inclinations
$\theta=5^\circ$ (thin solid lines), $60^\circ$ (medium lines), and 
$85^\circ$ (thick lines). The three panels correspond to the three
models as indicated. The flux units are arbitrary.
}
\end{figure}

The line profiles are obtained with the help of the same calculation
scheme, but the volume emissivity of matter given in Eq.(22) is replaced
by Eq.(24). The lines are appreciably broadened. 
The mechanism of line formation which we use (spallation reactions
leading to the production of lithium and beryllium nuclei in excited
state), gives the lines which are broadened by thermal motion of the hot
gas. The effects in light propagation (Doppler and gravitational shifts)
are superimposed on the natural line broadening. As can be seen on the
plots, these combined effects produce profiles without the
characteristic two-peak structure as in the case of Keplerian disk
(Hameury, Marck \& Pelat 1994). 
Such two-peak structure is present in the spectra of the
maximal surface brightness rings (compare Fig.3), but the light from
these parts of the sources cannot be observed separately by a distant
observer. Superposition of the spectra from all the rings removes the
effect, but some traces of it can be seen in the spectra of our $a=0$
and to lesser extent $a=0.5$ models. 

\section{Discussion and conclusions}

Our system of equations describing the relativistic advection dominated
accretion flow onto a Kerr black hole is based on the ACGL and ALP papers
but there are few differences. We use all the terms related 
to the radial and azimuthal motion of matter in the expression for shear,
which is responsible for heat dissipation. We also combine the equations 
for energy and angular momentum conservation to get the equation for the 
angular momentum gradient, which does not contain gradients of radial 
velocity. (ACGL neglect such terms). We neglect the viscous forces in the 
equation for radial momentum balance; this approximation is also used 
by ACGL). We take into account the contribution 
of the thermal energy to the total energy density introducing the specific
enthalpy $\mu$ into our equations. Our description is however limited to 
the equation of state for an ideal, nonrelativistic gas. This approximation
is reasonable, since the thermal energies of ions and electrons 
do not exceed their rest energies. The contribution to pressure 
of the magnetic field is low (5\%) and does not limit the validity of the
ideal gas approximation.

The most serious problem in our approach (and in other descriptions of ADAF
in the frame of vertically averaged disk structure) is the fact, that the
flow is not limited to the close vicinity of the equatorial plane. 
We obtain the  three dimensional description of the flow assuming that 
the flow is quasi-spherical (we neglect the component 
$u^\theta$ of the velocity) and that the density falls in the vertical
direction as in isothermal atmosphere, with a constant scale height
depending only on radius.

The influence of the relativistic effects in light propagation
on the shape of the continuum spectra of advection dominated accretion 
flows seems to be marginal. This is due to the fact that the emitted 
spectra are very broad and the small frequency shifts are hard to
distinguish. 

The most important effect we find is the anisotropy of the radiation
emitted from the disk, which favors observers near the equatorial plane.
Observers close to this plane can see part of the matter as
approaching them with the relativistic speed due to the combined effects of
rotation and fast radial fall. This effect can be stronger that the
occurrence of hot spots on the surface of thin Keplerian disk in the
places approaching the observer with the highest velocity (Luminet 1979;
Jaroszy\'nski, Wambsganss \& Paczy\'nski 1992). 
Since matter is transparent one can see the beaming of radiation 
regardless of the place of its emission. The fact, that the effect is
stronger in our models with slowly rotating holes is due to the
different angular momentum distributions in the accreting matter. 
Also the fact, that the flow on the fast rotating hole ($a=0.9$) has
a higher ion temperature (see the plots of the sound speed on Fig.1)
has an impact. Since the electron temperature does not grow with ion
temperature indefinitely but saturates at about $10^{10}$K, it makes the
central, fast rotating part of the flow relatively less important as a
source of radiation and relativistic effects become weaker than usually
expected for a fast rotating hole.

The shape of the hypothetical $\gamma$-ray line, which we calculate,
does not have a two-peak structure due to the superposition of photons coming
from the different parts of the flow. The specific production mechanism
of the line, which we use, also acts as to make the relativistic effects
less pronounced. The crossection for the lithium and beryllium production
decreases with the ion temperature. The hotter flow on the Kerr hole
with $a=0.9$ produces a stronger line, since the volume with the
temperature above the threshold is larger, but in the same time, the
central region is much less important.

In the conclusion we state, that the kinematics of advectively 
dominated accretion
flows is difficult to investigate on the basis of spectral
observations, since the possible pronounced spectral features are lacking. 
Also the parameter space of the possible ADAFs is much larger than in
the case of Keplerian disks, which makes the task even harder.

\begin{acknowledgements}
We thank Jean-Pierre Lasota and Marek Abramowicz 
for the discussions concerning ADAFs, and one of us (MJ) for their kind
hospitality during his visits to Meudon and G{\"o}theborg.
  This work was supported in part by the Polish State Committee
for Scientific Research grant 2-P03D-020-08.
\end{acknowledgements}

\end{document}